# Electron-overdoped Ag(II)F$_2$: mixed-valence fluorides Ag(I)Ag(II)F$_3$ and Ag(I)$_2$Ag(II)F$_4$

Katarzyna Kuder[a], Kacper Koteras[a], Zoran Mazej[b]* and Wojciech Grochala[a]*

This work is dedicated to Professor Marius Andruh at his 70$^{th}$ birthday

[a]  Prof. W. Grochala
     Center of New Technologies
     University of Warsaw
     Żwirki i Wigury 93, 02-089 Warsaw Poland
     E-mail: w.grochala@cent.uw.edu.pl
     Katarzyna Kuder, Dr Kacper Koteras
     Center of New Technologies
     University of Warsaw
     Żwirki i Wigury 93, 02-089 Warsaw Poland
[b]  Dr. Z. Mazej
     Department of Inorganic Chemistry and Technology
     Jožef Stefan Institute
     Jamova cesta 39, 1000 Ljubljana, Slovenia
     E-mail: zoran.mazej@ijs.si

**Abstract:** We have successfully prepared two novel mixed-valence compounds of silver, Ag(I)Ag(II)F$_3$ and Ag(I)$_2$Ag(II)F$_4$. They may be considered to be long-sought strongly electron-overdoped Ag(II)F$_2$. Their crystal structures indicate that both belong to the Class I (mixed-valence) family with frozen Ag(I) and Ag(II) valences. The measured Raman spectra are well-correlated with the theoretical ones. Density functional theory calculations reveal their smaller fundamental band gaps as compared to pristine AgF$_2$, due to the presence of Ag(I) states in the valence band.

## Introduction

Of the millions of known inorganic compounds, only a tiny fraction exhibits a formally non-integer oxidation state of an element. This occurs either in either mixed- or intermediate-valence systems [1,2]. Fe$_3$O$_4$ – magnetite – is an example of the first family, Ag$_2$F – silver subfluoride – of the second. Although the chemical analysis of magnetite shows that its chemical formula is "FeO$_{1.33}$", the oxidation state of the iron is not +2.66; in fact, structural analysis shows that this compound has both Fe(II) and Fe(III) in a 1:2 ratio, and thus its correct formula is Fe(II)Fe(III)$_2$O$_4$. At ambient conditions (p,T) the iron valences are "frozen" and electrons cannot readily be exchanged between the two. This is an example of a so called "Class I" or **mixed-valence** (MV) system in the Robin and Day's classification [1a]. The Prussian blue, i.e. Fe$_7$(CN)$_{18}$ (=Fe(III)$_4$[Fe(II)(CN)$_6$]$_3$), InS (=In(I)In(III)S$_2$), AgO (=Ag(I)Ag(III)O$_2$), AuCl$_2$ (=Au(I)Au(III)Cl$_4$), Sb$_2$O$_4$ (=Sb(III)Sb(V)O$_4$), BaBiO$_3$ (=Ba$_2$Bi(III)Bi(V)O$_6$], and many other systems, fall into this category. Regardless of whether Class I systems are molecular or extended solids, they are usually electrically insulating, and optical inter-valence charge-transfer (IVCT) excitation usually occurs in the UV or vis region of the electromagnetic spectrum. Therefore, thermal energy cannot overcome the barrier for electron transfer between the two metal centers (Figure 1).

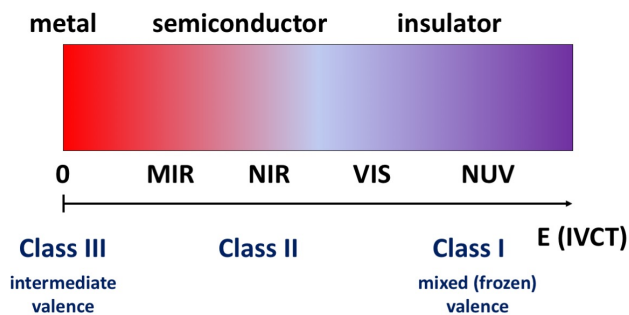

**Figure 1.** Classification of mixed-valence systems to genuine mixed-valence (Class I), intermediate valence (Class III) and intermediate class II, together with the associated electronic conductivity and energy of the optical IVCT band for extended solids.

On the other hand, Ag$_2$F is a genuine compound of Ag(+½) (i.e., **intermediate valence** (IV) between 0 and 1, or a Class III system [1a]), since all silver centers have the same coordination sphere. Importantly, when such systems form extended solids, they are usually metallic, as the bands are partially filled. In other words, there is no barrier to electron transfer between such metal centers. Consequently, the IVCT band is not present in their spectrum.

Class II systems [2a], such as the famous Creutz-Taube ion, {[Ru(NH$_3$)$_5$]$_2$(pyrazine)}$^{5+}$, have properties that lie between these two groups. Here the IVCT band is found in the NIR region of the spectrum, and in some systems of this type it can even fall into the MIR region. Obviously, at sufficiently high T, electronic transport can be activated, leading to a similarity of Class II systems to semiconductors (Figure 1).

All MV and IV systems, however rare they may be, are of immense importance for the understanding of electron transfer, – i.e. the fundamental process of chemistry. Moreover, some IV systems are among the most useful or fascinating materials for mankind. Think of Li$_x$CoO$_2$ (the classical ionic and electronic conductor in lithium ion batteries), La$_{1-x}$Ba$_x$CuO$_4$ (the first oxocuprate superconductor), LaH$_{10+x}$ (the first superconductor at high pressure near room temperature) or La$_{1-x}$Ca$_x$MnO$_3$ (the prototype of a giant magnetoresistive material), (p-C$_2$H$_2$)(I$_3$)$_x$ (the classical electronic conducting polymer), H$_x$WO$_3$ (the tungsten bronze) and many others.

The fluorides of silver(II) have been intensively studied over the last two decades due to numerous similarities between these materials and copper(II) oxides [3]. It is believed that AgF$_2$ is a close analogue of La$_2$CuO$_4$ [3b]. One of the most important prerequisites for achieving high-temperature superconductivity in these materials is that the Ag(II) system should be electron- or hole-doped in a similar way to copper(II) oxides. Indeed, in addition to stoichiometric AgF, AgF$_2$ and AgF$_3$, three mixed-valence silver fluorides are known: Ag$_2$F [4a], Ag$_2$F$_5$ [4b] and Ag$_3$F$_8$ [4c] (Figure 2). While the former is an IV system, the latter two are of the MV type. Ag$_2$F$_5$ and Ag$_3$F$_8$ can be considered as strongly hole (h$^+$)-overdoped AgF$_2$ derivatives. However, their electron (e$^-$)-doped analogues have been lacking so far. Their preparation and characterization are the main goals of this work.

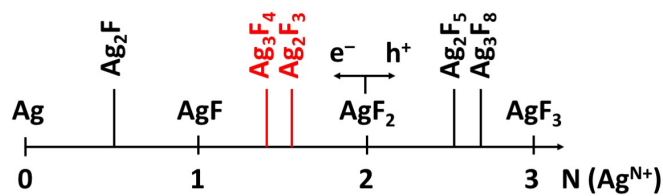

**Figure 2.** The known stoichiometric (single valence) and MV or IV fluorides of silver, with the two new stoichiometries targeted in this work (in red); N is the formal oxidation state of silver.

## Results and Discussion

### Synthesis

We applied a classical solid-solid thermal synthesis using AgF$_2$ and AgF for the preparation of the title compounds. Since the AgF$_2$ substrate is highly oxidizing and fluorinating we used a nickel reactor (nickel is easily passivated). We tested different temperature conditions, but we found that Ag$_3$F$_4$ repeatedly formed at T ~ 340 °C. On the other hand, we succeeded to prepare Ag$_2$F$_3$ only once, starting from the same reagent molar ratio of 2:1 (see section S1 in SI for the synthesis protocol). Unfortunately, all attempts to repeat this synthesis failed, usually resulting in a mixture of AgF$_2$ and Ag$_3$F$_4$. This fact alone suggests that Ag$_2$F$_3$ is a very shallow minimum on the potential energy surface (PES) and an inherently metastable system. It appears that the the reaction (Eq.1) is the key channel for its thermal instability:

2 Ag$_2$F$_3$ → AgF$_2$ + Ag$_3$F$_4$     (Eq.1)

Previous theoretical calculations have indeed indicated that this reaction is not far from equilibrium, with calculated reaction energies in the order of –10 meV per mole of Ag$_2$F$_3$ [5]. This means that a small change in the experimental parameters can push the system with such a stoichiometry (thermodynamically) out of the desired minimum. Interestingly, very similar observations on the metastability of Ag$_2$F$_3$ were recently reported in a not-yet-peer-reviewed manuscript deposited on a preprint server [6]. These authors succeeded in preparing Ag$_2$F$_3$ more reproducibly by mechanochemical synthesis in combination with cooling of the milling vessel in LN2. However, as we will see, these two approaches lead to two different polymorphic forms.

### Crystal structure of Ag$_2$F$_3$

We were able to successfully model the measured powder X-ray pattern (see section S2 of SI for technical details and section S3 for crystallographic tables) using a two-phase model (Figure 3). The only impurity phase detected was AgF (at 55.08(99) wt. %); no traces of AgF$_2$ were detected. The second phase in this sample was determined to be the new I MV fluoride Ag$_2$F$_3$. This corresponds to the report in [6], in which Ag$_2$F$_3$ (approx. 57.4 wt. %), which was significantly contaminated with AgF$_2$ and Ag$_3$F$_4$, was prepared.

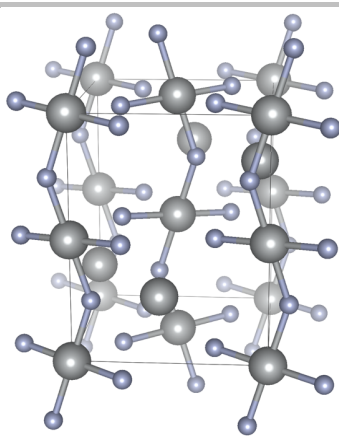

**Figure 3.** The crystallographic unit cell of the orthorhombic form of $Ag_2F_3$, highlighting the presence of $[AgF_{2/1+2/2}]^-$ bent 1D chains.

AgAgF$_3$ adopts an orthorhombic $P$nma crystal structure [7] (KAgF$_3$-type [8]) with unit cell parameters of 6.5360(10) Å, 7.5574(11) Å, and 5.8458(6) Å (see Table S1 in the Supplementary Information). Interestingly, our structure differs from that reported in [6], where a triclinic structure with much lower symmetry of the AgCuF$_3$-type was obtained, regardless of whether mechanochemical or a solid-state synthesis was used. Our phase also differs from the theoretically predicted ground state of the monoclinic CaIrO$_3$-type [5b]. Comparison of the volumes per four formula units for the triclinic cell (295.58(1) Å$^3$) [6] and our orthorhombic cell (288.75(7) Å$^3$) suggests that the polymorph obtained here is a low-temperature form, whereas the form prepared in [6] is a high-temperature form. This hypothesis seems to contradict with the fact that one of the synthesis methods used in [6] involved LN2 cooling. However, a peculiarity of mechanochemical synthesis is that an enormous amount of kinetic energy is transferred to the small amount of reacting powders; the local temperature at the surface of the grains can be extremely high, while LN2 merely serves as a quenching factor. The fact that the LT polymorph has a higher crystallographic symmetry than the HT polymorph is also quite unusual; more often an increase in T leads to a higher symmetry. Although our polymorph is quite elusive, it still provides valuable information about the unusually complex PES in the AgF/AgF$_2$ phase diagram.

The crystal structure of orthorhombic Ag$_2$F$_3$ exhibits the features known for the related distorted perovskite, KAgF$_3$ [8], i.e. the presence of the infinite [Ag)II)F$^+$] chain along the crystallographic **b**-direction and the antiferrodistortive Ag(II)–F...Ag(II) bonding pattern within the [Ag(II)F$_2$] sheets. The Ag(II)–F bond lengths are 1.983(15) Å (x2) and 1.93(5) Å (x2) with two longer distances of 2.60(5) Å. The first coordination sphere of Ag(I) is quite irregular. We note that KAgF$_3$ exhibits a structural transition related to orbital-rearrangement [9]; to verify whether similar transition (or possibly some substitutional disorder at F sites) is present in Ag$_2$F$_3$ the positions of the light atom positions would need to be precisely determined. The subtleties of the positions of the light atom and a possible disorder can be investigated in the future, e.g. with neutron diffraction [9].

## Lattice dynamics of Ag$_2$F$_3$

Figure 4 shows the Raman spectrum of Ag$_2$F$_3$, which was recorded with an excitation of 532 nm. The spectrum is dominated by the very strong 416 cm$^{-1}$ feature with a handful of weaker side bands (Table 1). We have observed a similar feature several times in the past when AgF$_2$ was irradiated with green laser light [10,3b]. This suggests that AgF$_2$ can be photochemically decomposed, simultaneously releasing F$_2$ gas to form Ag$_2$F$_3$ (and/or the related Ag$_3$F$_4$; we will return to this later).

To understand the spectral features, we have performed density functional theory (DFT) calculations of the phonon spectra (see section S4 in SI). The theory yields no imaginary modes for the orthorhombic structure of Ag$_2$F$_3$ confirming that it is a local minimum on the PES. Furthermore, we were able to assign the bands appearing in the Raman spectrum, four of which are from vibrational fundamental modes and three from combination modes (Table 1). The agreement between experimental and theoretical values is very good without scaling factor. The complete list of modes predicted by the DFT can be found in SI (section S4).

The fact that a 416-420 cm$^{-1}$ mode which is not totally-symmetric is the strongest in the spectrum and that its overtones do not appear suggests that the spectrum does not have a strong resonance-Raman character. Nevertheless, the two high-wavenumber combination modes also contain the counterpart at 416-420 cm$^{-1}$, i.e. they "borrow intensity" from the strongest mode mentioned. This means that the 532 nm (2.33 eV) laser line is in weak resonance with some electronic transition. We note that the spectrum measured here differs from that reported in [6] for another Ag$_2$F$_3$ polymorph. This is normal because a vibrational spectrum is a fingerprint of any chemical compound.

It is also interesting to note that the recently reported triclinic structure [6], for which the calculated energy at 0K is 7.7 meV larger than for our structure, has significant dynamic (phonon) instabilities (see SI, section S4); this suggests that the triclinic structure may be incorrect and the actual structure either has some disorder or is a superstructure of what was reported.

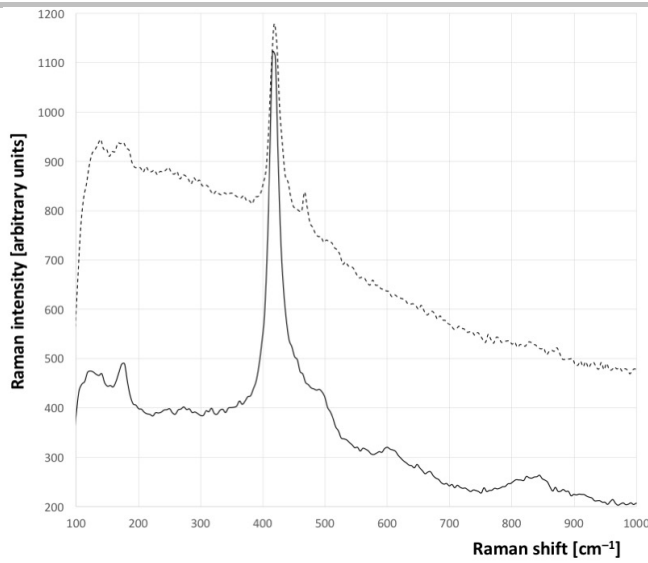

**Figure 4.** Raman spectrum of $Ag_2F_3$ (bottom) and $Ag_3F_4$ (top).

**Table 1.** Wavenumbers (ν) of bands appearing in the Raman spectra of $Ag_2F_3$ and $Ag_3F_4$ together with their theoretical assignment (symmetry labels are given); w – weak, m – medium, s – strong, v – very, sh – shoulder.

| | $Ag_2F_3$ | | | $Ag_3F_4$ | |
|---|---|---|---|---|---|
| $v_{exp}$ [cm$^{-1}$] | $v_{theor}$ [cm$^{-1}$] | symmetry | $v_{exp}$ [cm$^{-1}$] | $v_{theor}$ [cm$^{-1}$] | symmetry |
| 122 (vw) | 122 | $B_{1g}$ | 139 (vw) | 127 | $A_g$ |
| 177 (vw) | 173 | $B_{2g}$ | 176 (vw) | 169 | $B_g$ |
| 416 (vs) | 426 | $B_{2g}$ | 418 (vs) | 407 | $B_g$ |
| 450 (sh) | 445 | $B_{3g}$ | 441 (sh) | 407+40 | $A_g$ |
| 495 (sh) | 426+67 | $A_g$ | 467 (w) | 407+65 | $B_g$ |
| 605 (vw) | 426+173 | $A_g$ | --- | | |
| 849 (w, broad) | 426+445 | $A_g$ | --- | | |

**Electronic and magnetic structure of $Ag_2F_3$**

Periodic DFT+U calculations were performed to gain insight into the electronic structure and magnetic properties of $Ag_2F_3$. We found that the lowest energy magnetic state is an AFM-G type antiferromagnetic state with ferromagnetically ordered [$AgF_2$] layers and antiferromagnetically ordered [AgF]$^+$ chains. This type of ordering is found in other Ag(II)/F post-perovskite materials [7c,8]. The predicted magnetic superexchange constants J are equal to –53.1 meV (intra-chain) and +0.1 meV (within [$AgF_2$] layers on *ac* planes). The absolute value of intra-chain J is about half of the t measured and calculated value for the $KAgF_3$ analogue [8a,9]. This is consistent with the Ag-F-Ag intrachain angle being more bent for $Ag_2F_3$ (145(3)°) than for $KAgF_3$ (153.2(7)° [8b] and 157.2(8)° [8a], respectively), i.e. in agreement with the Goodenough-Kanamori rules [11]. Since the value of the said angle for the triclinic form (149.5(7)° [6]) is similar to the value obtained here for the orthorhombic form, we conclude that the intra-chain J values should also be similar in both forms. Therefore, the rather low order point of the triclinic point, $T_N$=7.3(7)K [6], seems to be due to the weak inter-chain interactions. As we have already mentioned, these interactions are also weak in the orthorhombic form and are of the order of 0.1 meV (= approx. 11.6 K).

The electronic band structure of orthorhombic $Ag_2F_3$ shows the fundamental band gap of 2.14 eV (see section S5 in SI). Since this type of calculation tends to underestimate the gap for $AgF_2$, we assume that the actual gap is larger (mainly because the laser line of 2.33 eV is in weak resonance with any electronic transition). However, it is important to note that the band gap calculated for $AgF_2$ using the same method is 2.44 eV. The narrowing of the band gap during formation of the mixed-valence system is to be expected since the filled Ag(I) states are positioned between the valence and conduction bands of $AgF_2$ (see section S4 in SI). That the narrowing of the gap actually occurs can be seen from the color of $Ag_2F_3$ which is black in contrast to $AgF_2$ (brown).

The gap falling into the visible or even UV region in $Ag_2F_3$ is consistent with the MV rather than IV nature of this system; the coordination spheres of Ag(II) and Ag(I) are very different, leading to large reorganization energy in electron transfer between the two.

Having described the elusive polymorphism of metastable $Ag_2F_3$ we now turn to $Ag_3F_4$.

### Crystal structure of $Ag_3F_4$

Using very similar synthetic conditions as described above, we succeeded in preparing $Ag_2AgF_4$. In contrast to the previous synthesis, this synthesis is quite repeatable and more tolerant to variations in reaction temperature.

The obtained specimen contains only some unreacted AgF (31.23(76) wt. %). The main phase is monoclinic $Ag_3F_4$, crystallizing in space group $P2_1/c$ with unit cell vectors 3.5604(6) Å, 9.8794(14) Å, 6.8063(11) Å, $\beta$ = 118.595(12)° (Figure 5 and Table S2 in the Supplementary Information). This turns out to be identical to that reported in [6] when transformed by a rotation matrix. Indeed, the volumes of the unit cells for our solution (210.21(6) Å$^3$) and the one reported in [6] (210.86(1) Å$^3$) are practically identical.

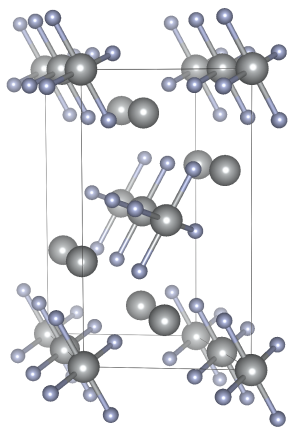

**Figure 5.** The crystallographic unit cell of the monoclinic form of $Ag_3F_4$ highlighting the presence of $[AgF_{4/1}]^{2-}$ square plaquettes stacked in 1D chains (DFT result).

The Ag(II)–F bond lengths are 2.52(7) Å (x2) and 2.58(7) Å (x2) with two shorter separations of 2.07(10) Å. The first coordination sphere of Ag(I) is quite irregular.

### Lattice dynamics of $Ag_3F_4$

The Raman spectrum recorded for our sample (Figure 4) is similar to the spectrum measured for [6]. This is to be expected since both structures are identical. On the other hand, it is surprising that the Raman spectrum recorded for $Ag_3F_4$ is generally similar to that reported here for $Ag_2F_3$. However, there are also some differences. For example, a sharp band at 467 cm$^{-1}$ is present in the spectrum of $Ag_3F_4$, whereas it is not present in $Ag_2F_3$. The band with the lowest measured wavenumber is at 122 cm$^{-1}$ for $Ag_2F_3$ but at 139 cm$^{-1}$ for $Ag_3F_4$ and so on. The bands originating from overtones also differ for the two compounds. While the spectrum measured here for $Ag_3F_4$, appears to be a fingerprint of this species, it cannot be ruled out that both compounds contain small amounts of the other on the surface of their grains. While Rietveld analysis does not reveal their presence due to the low sensitivity of this method (about 5 wt. %), Raman spectra measured with a confocal Raman microscope and focused on the surface of a small grain may be affected.

We tried to assign the bands appearing in the spectra by quantum mechanical calculations of the phonon modes. To our surprise, the phonon dispersion shows several imaginary modes (see SI, section S4). This suggests that the structure reported here and in [6] may not be the true structure of $Ag_3F_4$. It is possible that the actual structure either has some disorder or is a superstructure of what has been reported. Unfortunately, it is not possible to calculate phonon spectra for disordered structures using a periodic model; moreover, the determination of possible disorder at F sites requires the use of neutron diffraction, since F atoms do not scatter the X-ray beam sufficiently strongly. Therefore, we are forced to postpone the solution of this problem to future investigations.

### Electronic and magnetic structure of $Ag_3F_4$

Since an ordered structure is a reasonable approximation to the postulated disordered model, we decided to calculate the electronic structure of $Ag_3F_4$ in the ordered unit cell reported in [6] and here (Figure 6). The lowest energy magnetic model for this structure exhibits AFM ordering along the chains and FM ordering between the chains.

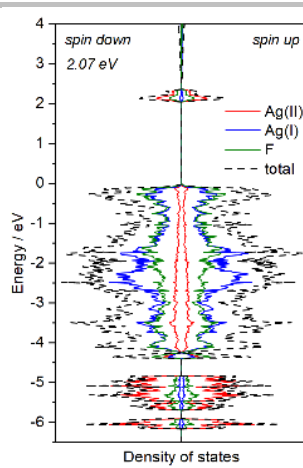

**Figure 6.** The electronic structure of the ordered monoclinic form of $Ag_3F_4$. The Ag(II), Ag(I) and F states are shown together with the total electronic DOS.

The electronic band structure shows the fundamental band gap of 2.07 eV, which is probably underestimated. The filled Ag(I) states together with F(2p) states form the valence band, while the conduction band corresponds to the upper Hubbard band and is dominated by Ag(II) states (Figure 6). The calculated SE constants are very small, below 0.5 meV, which is consistent with the quasi-0D electronic and magnetic nature of the compound with isolated [Ag(II)$F_4$] squares.

## Conclusion

Mixed valence systems continue to fascinate physicists, chemists and materials scientists. Here we have prepared and preliminarily characterized two new systems, namely Ag(I)Ag(II)$F_3$ and Ag(I)$_2$Ag(II)$F_4$. Both are mixed- rather than intermediate-valence substances, with distinct crystallographic positions of Ag(I) and Ag(II). These substances are black in color, indicating a rather narrow fundamental band gap of about 2–3 eV, suggesting their electrically insulating character. According to the DFT calculations, both should exhibit magnetic order at sufficiently low temperatures. The bands appearing in their Raman spectra were assigned to diverse phonon modes based on QM calculations. Although it appears that two new chemical systems have been successfully produced, they still hold their secrets.

First, $Ag_2F_3$ is extremely unstable or rather metastable with respect to an equimolar mixture of $AgF_2$ and $Ag_3F_4$. This is reminiscent of the classical "disappearing polymorphs" as beautifully described in a seminal 1995 paper by Dunitz and Bernstein [12]. Moreover, the crystal structure described here is different from that found in [6], and both polymorphs are of the "disappearing" sort…! Further studies are needed to understand how both polymorphs can be repeatedly prepared in pure form.

Secondly, although the $Ag_3F_4$ described here is identical to that discussed in [6], and both correspond to the long-predicted $Na_2AgF_4$-type [5a], its phonon dispersion suggests that the ordered crystal structure is only a crude approximation to reality. The correct structure will either have some static disorder and/or be a superstructure of the reported structures, with small differences in atomic position within each of the original unit cells. It is hoped that these issues can be clarified by future investigations.

## Supporting Information

SI contains experimental and computational details, as well as selected supplementary results. The authors have cited additional references within the Supporting Information.[13-20]

## Acknowledgements

This research was supported by the Polish National Science Center (NCN) within WG's Maestro project (2017/26/A/ST5/00570). The research was carried out with the use of CePT infrastructure financed by the European Union – the European Regional Development Fund within the Operational Programme "Innovative economy" for 2007–2013 (POIG.02.02.00-14-024/08-00). ZM acknowledges the financial support of the Slovenian Research and Innovation Agency (research core funding No. P1-0045; Inorganic Chemistry and Technology). Calculations have been carried out using resources provided by the Interdisciplinary Centre of Mathematical and Computational Modelling (grant SAPPHIRE GA83-34).

# SUPPLEMENTARY INFORMATION

## Electron-overdoped Ag(II)F$_2$: mixed-valence fluorides Ag(I)Ag(II)F$_3$ and Ag(I)$_2$Ag(II)F$_4$

Katarzyna Kuder[a], Kacper Koteras[a], Zoran Mazej[b]* and Wojciech Grochala[a]*

[a] Center of New Technologies, University of Warsaw, Zwirki i Wigury 93, 02-089 Warsaw Poland

E-mail: w.grochala@cent.uw.edu.pl

[b] Department of Inorganic Chemistry and Technology, Jožef Stefan Institute, Jamova cesta 39, 1000 Ljubljana, Slovenia

E-mail: zoran.mazej@ijs.si

## S1.   Synthesis

All manipulations were performed under anhydrous conditions. Non-volatile substances were handled in a M. Braun glove box in an argon atmosphere in which the moisture content did not exceed 0.5 ppm. Volatile compounds, such as aHF and F$_2$, were handled in a vacuum line constructed from nickel and PTFE (polytetrafluoroethylene) parts. The vessel used for the fluorination of AgF at room temperature consisted of a tube (i.d. 16 mm, o.d. 19 mm) made of tetrafluoroethylene-hexafluoropropylene block-copolymer (FEP; Polytetra GmbH, Germany). It was heat-sealed at one end and fitted with a PTFE valve at the other end.

Commercially available AgF (Aldrich, 99.9%) was treated with a small amount of elemental fluorine in anhydrous HF at room temperature. When the first traces of brown AgF$_2$ began to precipitate, the remaining AgF/HF solution was decanted and the volatiles pumped off. In this way, fresh AgF was prepared without traces of metallic Ag. The AgF$_2$ was prepared by fluorination of AgNO$_3$ (Alfa Aesar, 99.9%) with elemental fluorine at 250 °.

**Ag$_3$F$_4$ (sample 1)** The initial mixture of 254 mg (2.0 mmol) AgF and 146 mg (1.0 mmol) AgF$_2$ was ground on an agate mortar, placed in a Pt-boat covered with a Pt-lid and sealed in a nickel reactor (130 ml) under a dry argon atmosphere. The reactor was evacuated. Under these conditions, the mixture of 2AgF / AgF$_2$ was heated to about 340 °C. After 25 hours of heating, the reactor was cooled to room temperature and the mixture was ground again. The synthesis procedure was repeated. After a further annealing at 350-358 °C for 47 hours, the light brown sample was obtained and analyzed.

**Ag$_2$F$_3$ (sample 2)** The second experiment between 130 mg (1.02 mmol) AgF and 175 mg (0.51 mmol) AgF$_2$ was carried out in a similar way, except that the mixture was annealed only once at 350 °C for 24 hours. The sample obtained appears darker, i.e. dark brown, than the product of the first reaction.

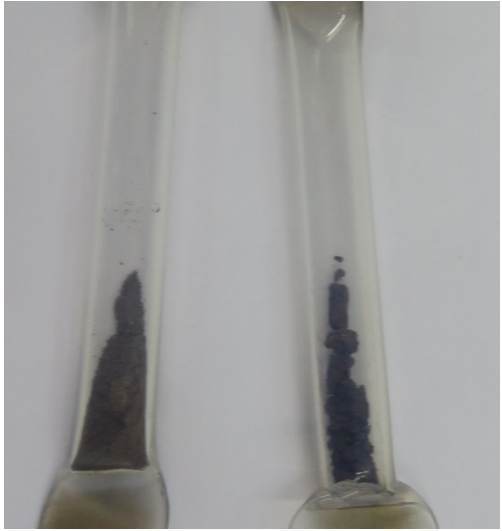

Figure 1. Sample 1 (left); sample 2 (right).

## S2. Other experimental and computational details

Powder X-ray diffraction experiments were conducted XPERT PRO diffractometer using CoKα radiation. Prefluorinated quartz capillary with the size of 0.5 mm was used in order to minimize absorption and decomposition.

Results were analysed in Fullprof software suite [13] by the means of Rietveld refinement. [14] With the use of that method, it is possible to fit unit cell parameters and atomic positions. Isotropic atomic displacement factors were set based on literature data.

All calculations were performed in VASP 5.4.4 suite using projector-augmented-wave method. [15, 16] Core electrons were described using standard VASP pseudopotentials. Strong valence electrons correlation was accounted for within the GGA+U framework using Lichtenstein formalism [17] ($U_{Ag}$ = 8.0 eV and $J_{Ag}$ = 1.0 eV [18]), and Perdew-Burke-Ernzerhof DFT-GGA functional revised for solids (PBEsol).

The cut-off energy of the plane wave basis set was equal to 520 eV, with k-point mesh spacing equal to 0.16 Å$^{-1}$. Geometry relaxations were conducted with default VASP algorithms, and the self-consistent-field convergence criterion was set to $1 \cdot 10^{-7}$ eV per unit cell. Whether atomic positions or unit cell volume were relaxed during is specified in the text. Symmetry was turned off at most times, except when noted otherwise.

Lattice dynamics calculations were conducted using finite differences method either implemented in VASP. This method relies upon calculations of total energy of a set of displaced structures. Afterwards the set of oscillator energies is calculated. When modelling phonons using DFT+U method, an appropriate supercell was used as a starting structure to increase precision and using energy cutoff of 800 eV. Phonon band structures were calculated for standardized k-path [19] in Phonopy software package. [20]

Magnetic properties were calculated using increased precision settings using denser grid used to represent pseudopotentials and increased energy cut-off of 800 eV. Magnetic superexchange constants for were calculated using the broken symmetry method mapping calculated energy values on Heisenberg Hamiltonian in a form:

$$H = E_0 - 0.5 \sum_{i,j} J_{ij} S_i S_j ,$$

where $E_0$ is the total energy of a non-magnetic state, $S$ and $S$ are unpaired electrons' spins and $J_{ij}$ is coupling constant between $i$th and $j$th electrons. Expanding that sum, with the inclusion of selected exchange pathways and with consideration of crystal symmetry, over all spins in a unit cell yields a total energy model. By calculating the total energy of different magnetic configurations, it is possible to derive the superexchange coupling constants.

## S3. Crystal structures

Table S1. Crystallographic data for the $Ag_2F_3$ phase.

| Formula | AgAgF$_3$ | | | | | | |
|---|---|---|---|---|---|---|---|
| Colour | black | | | | | | |
| Space group | *Pnma* (No. 62) | | *x* | *y* | *z* | *Occ.* | $U_{iso}$ |
| Z | 4 | | | | | | |
| V / Å$^3$ | 288.75(7) | Ag1 | 0.50000 | 0.00000 | 0.50000 | 1.0 | 0.00633 |
| a / Å | 6.5360(10) | Ag2 | 0.0946(16) | 0.25000 | 0.4400(16) | 1.0 | 0.00633 |
| b / Å | 7.5574(11) | F1 | 0.517(10) | 0.25000 | 0.399(8) | 1.0 | 0.01267 |
| c / Å | 5.8458(6) | F2 | 0.344(7) | 0.521(7) | 0.221(8) | 1.0 | 0.01267 |
| α / ° | 90 | | | | | | |
| β / ° | 90 | R$_P$ = 1.44 % | | | cR$_P$ = 53.72 % | | |
| γ / ° | 90 | R$_{wp}$ = 2.14 % | | | cR$_{wp}$ = 26.74 % | | |

Table S2. Crystallographic data for the $Ag_3F_4$ phase.

| Formula | Ag$_2$AgF$_4$ | | | | | | |
|---|---|---|---|---|---|---|---|
| Colour | black | | | | | | |
| Space group | *P*2$_1$/*c* (No. 12) | | *x* | *y* | *z* | *Occ.* | $U_{iso}$ |
| Z | 2 | | | | | | |
| V / Å$^3$ | 210.21(6) | Ag1 | 0.00000 | 0.00000 | 0.00000 | 1.0 | 0.01900 |
| a / Å | 3.5604(6) | Ag2 | 0.213(3) | 0.8386(13) | 0.4534(15) | 1.0 | 0.01900 |
| b / Å | 9.8794(14) | F1 | -0.21(2) | 0.453(9) | 0.727(13) | 1.0 | 0.04813 |
| c / Å | 6.8063(11) | F2 | 0.444(18) | 0.164(9) | 0.712(12) | 1.0 | 0.04813 |
| α / ° | 90 | | | | | | |
| β / ° | 118.595(12) | R$_P$ = 1.74 % | | | cR$_P$ = 87.6 % | | |
| γ / ° | 90 | R$_{wp}$ = 2.80 % | | | cR$_{wp}$ = 48.2 % | | |

## S4. Lattice dynamics calculations

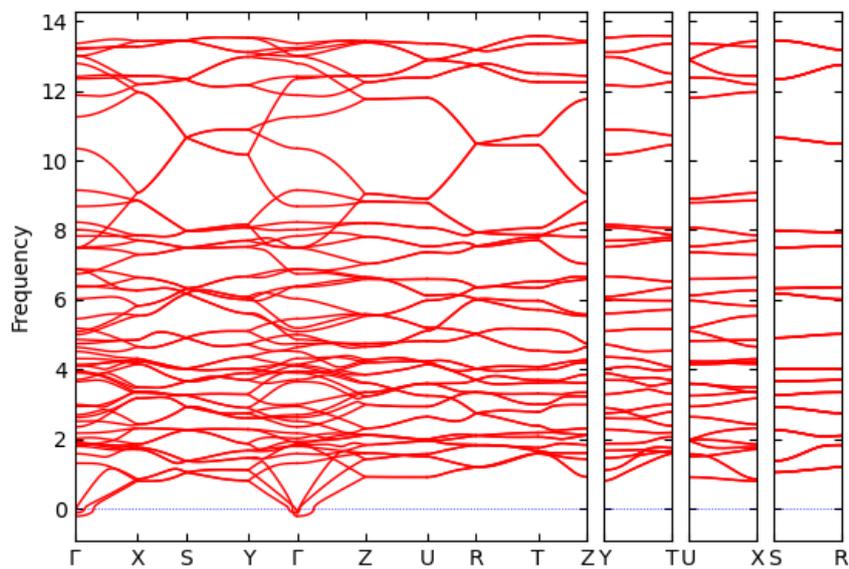

Figure S1. Phonon dispersion curves calculated for $Ag_2F_3$ phase (our orthorhombic model).

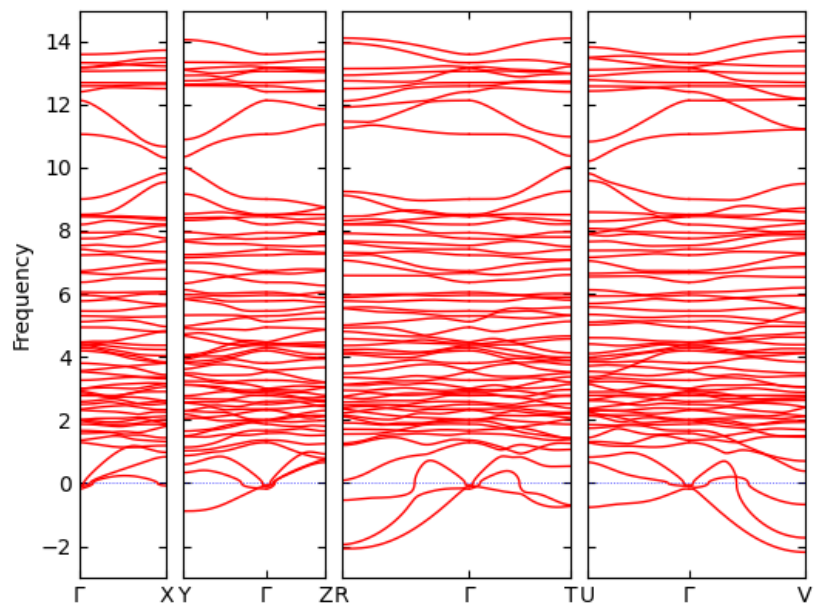

Figure S2. Phonon dispersion curves calculated for $Ag_2F_3$ phase (triclinic model published in [6]).

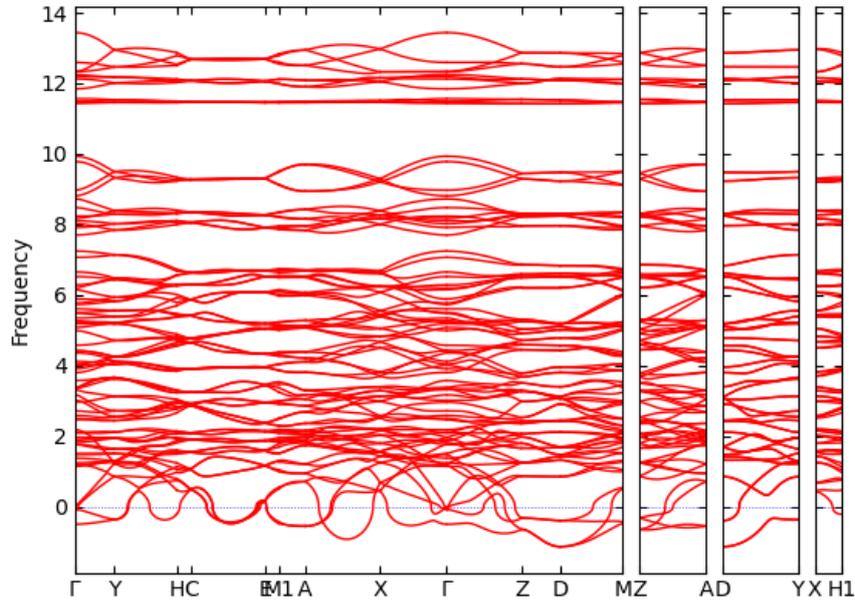

Figure S3. Phonon dispersion curves calculated for $Ag_3F_4$ phase.

Table S1. Calculated modes list for both phases.

| $Ag_2F_3$ modes | | | | $Ag_3F_4$ modes | | | |
|---|---|---|---|---|---|---|---|
| No | Wavenumber (THz) | Wavenumber (cm$^{-1}$) | Symm. | No | Wavenumber (THz) | Wavenumber (cm$^{-1}$) | Symm. |
| 1 | -0.210 | -7.0 | B2u | 1 | -0.503 | -17 | Au |
| 2 | -0.110 | -3.7 | B1u | 2 | -0.095 | -3 | Bu |
| 3 | -0.041 | -1.4 | B3u | 3 | -0.061 | -2 | Bu |
| 4 | 1.313 | 43.8 | Au | 4 | -0.014 | 0 | Au |
| 5 | 1.586 | 52.9 | B3u | 5 | 1.188 | 40 | Bg |
| 6 | 1.799 | 60.0 | B3g | 6 | 1.196 | 40 | Ag |
| 7 | 1.837 | 61.3 | Ag | 7 | 1.373 | 46 | Ag |
| 8 | 1.867 | 62.3 | B2g | 8 | 1.956 | 65 | Ag |
| 9 | 1.887 | 62.9 | Au | 9 | 1.963 | 65 | Bu |
| 10 | 1.937 | 64.6 | B2u | 10 | 2.175 | 73 | Bg |
| 11 | 2.023 | 67.5 | B1g | 11 | 2.463 | 82 | Bu |
| 12 | 2.174 | 72.5 | B1u | 12 | 2.496 | 83 | Au |
| 13 | 2.353 | 78.5 | B2u | 13 | 2.712 | 90 | Ag |
| 14 | 2.513 | 83.8 | Ag | 14 | 3.168 | 106 | Bu |
| 15 | 2.626 | 87.6 | Au | 15 | 3.363 | 112 | Au |
| 16 | 2.699 | 90.0 | B1u | 16 | 3.373 | 113 | Bg |
| 17 | 2.945 | 98.2 | B3u | 17 | 3.800 | 127 | Ag |
| 18 | 2.994 | 99.9 | B2g | 18 | 3.909 | 130 | Bu |

| | | | | | | |
|---|---|---|---|---|---|---|
| 19 | 3.649 | 121.7 | B1g | 19 | 4.358 | 145 | Bg |
| 20 | 3.708 | 123.7 | Ag | 20 | 4.420 | 147 | Au |
| 21 | 3.904 | 130.2 | B2u | 21 | 4.573 | 153 | Ag |
| 22 | 3.940 | 131.4 | Au | 22 | 4.990 | 166 | Bu |
| 23 | 3.953 | 131.9 | B1u | 23 | 5.077 | 169 | Bg |
| 24 | 4.103 | 136.9 | B2g | 24 | 5.215 | 174 | Au |
| 25 | 4.120 | 137.4 | B3u | 25 | 5.344 | 178 | Bu |
| 26 | 4.144 | 138.2 | B3u | 26 | 5.479 | 183 | Au |
| 27 | 4.315 | 143.9 | B3g | 27 | 5.728 | 191 | Ag |
| 28 | 4.520 | 150.8 | B1u | 28 | 5.884 | 196 | Bg |
| 29 | 4.637 | 154.7 | Ag | 29 | 6.285 | 210 | Au |
| 30 | 4.741 | 158.1 | Au | 30 | 6.652 | 222 | Bu |
| 31 | 4.865 | 162.3 | B2u | 31 | 7.698 | 257 | Ag |
| 32 | 4.999 | 166.7 | B3u | 32 | 8.018 | 267 | Bu |
| 33 | 5.087 | 169.7 | B1g | 33 | 8.169 | 272 | Bg |
| 34 | 5.186 | 173.0 | B2g | 34 | 8.720 | 291 | Au |
| 35 | 5.455 | 182.0 | B1u | 35 | 8.971 | 299 | Ag |
| 36 | 6.027 | 201.0 | B3g | 36 | 9.931 | 331 | Bg |
| 37 | 6.367 | 212.4 | B3u | 37 | 11.429 | 381 | Bu |
| 38 | 6.397 | 213.4 | B2u | 38 | 11.461 | 382 | Au |
| 39 | 6.744 | 225.0 | Ag | 39 | 12.189 | 407 | Bg |
| 40 | 6.881 | 229.5 | B1u | 40 | 12.250 | 409 | Au |
| 41 | 6.886 | 229.7 | Au | 41 | 12.331 | 411 | Au |
| 42 | 7.482 | 249.6 | B2g | 42 | 13.436 | 448 | Bu |
| 43 | 7.489 | 249.8 | B3g | | | | |
| 44 | 7.507 | 250.4 | Ag | | | | |
| 45 | 7.828 | 261.1 | B3u | | | | |
| 46 | 8.012 | 267.3 | B1u | | | | |
| 47 | 8.227 | 274.4 | B2g | | | | |
| 48 | 8.680 | 289.5 | B1u | | | | |
| 49 | 9.144 | 305.0 | B3u | | | | |
| 50 | 10.340 | 344.9 | B1g | | | | |
| 51 | 11.248 | 375.2 | Ag | | | | |
| 52 | 11.871 | 396.0 | B1g | | | | |
| 53 | 12.341 | 411.6 | Au | | | | |
| 54 | 12.407 | 413.8 | B2u | | | | |
| 55 | 12.782 | 426.4 | B2g | | | | |
| 56 | 12.985 | 433.1 | B2u | | | | |
| 57 | 13.003 | 433.7 | Au | | | | |
| 58 | 13.182 | 439.7 | B1u | | | | |
| 59 | 13.227 | 441.2 | B3u | | | | |
| 60 | 13.355 | 445.5 | B3g | | | | |

## S5. Electronic structure calculations

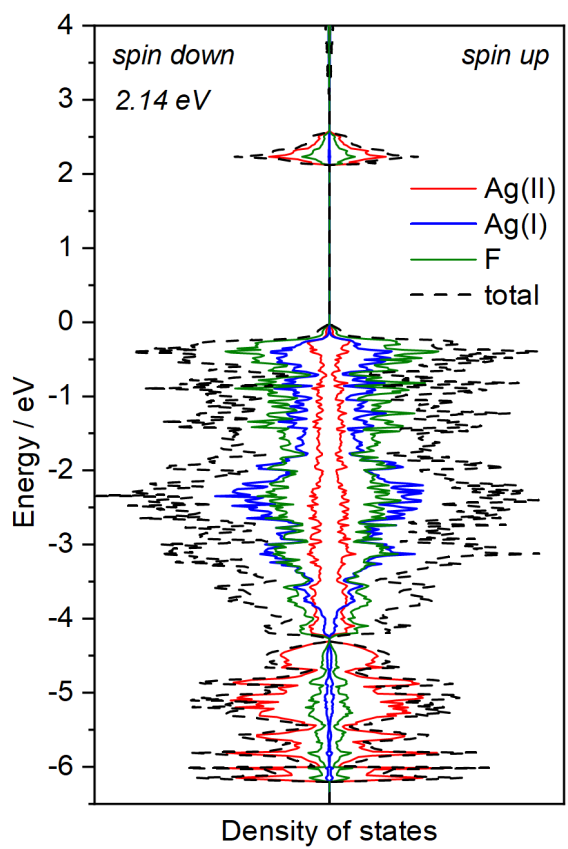

Figure S4. Projected electronic density of states graph for Ag$_2$F$_3$ phase.

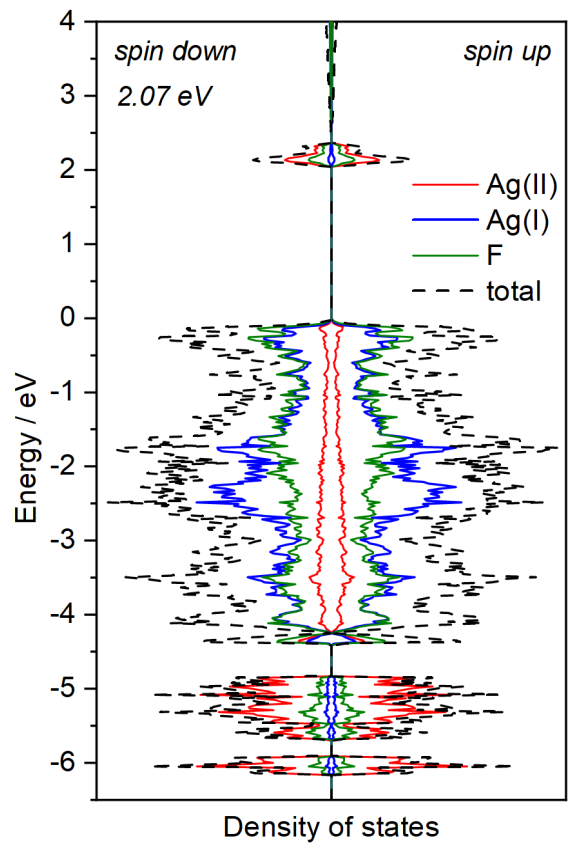

Figure S5. Projected electronic density of states graph for Ag$_3$F$_4$ phase.